\documentclass[prl, twocolumn, floatfix]{revtex4-2}
\setcitestyle{super}

\usepackage{graphicx}
\usepackage{hyperref}
\usepackage{amsmath}
\usepackage{color}
\usepackage{calc}

\begin{document}

\title{Spontaneously sliding multipole spin density waves in cold atoms}

\author{G.\ Labeyrie$^{1}$, J. G. M. Walker$^{2}$, G. R. M. Robb$^{2}$, R.\ Kaiser$^{1}$, and T.\ Ackemann$^{2}$\footnote{To whom correspondence should be addressed.}}
\affiliation{$^{1}$Universit\'{e} C\^{o}te d'Azur, CNRS, Institut de Physique de Nice, 06560 Valbonne, France}
\affiliation{$^{2}$SUPA and Department of Physics, University of Strathclyde,Glasgow G4 0NG, Scotland, UK}

\begin{abstract}
We report on the observation of spontaneously drifting coupled spin and quadrupolar density waves in the ground state of laser driven Rubidium atoms. These laser-cooled atomic ensembles exhibit spontaneous magnetism via light mediated interactions when submitted to optical feedback by a retro-reflecting mirror. Drift direction and chirality of the waves arise from spontaneous symmetry breaking.  The observations demonstrate a novel transport process in out-of-equilibrium magnetic systems. 

\end{abstract}

\maketitle


Magnetic properties of materials have been under intense scrutiny for decades, including complex and yet not fully understood phenomena as its connection to high-T$_c$ superconductivity~\cite{zhou21}. Exotic magnetic properties associated to high-order multipole states (quadrupole and beyond) have also recently attracted some interest in heavy-fermion metals~\cite{onimaru05,dabrowski14,jang17,tazai19,hafner22}, not the least due to the connection to unconventional superconductivity. Ref.~\cite{hafner22} concludes on the possibility of  the  existence of a quadrupolar density wave (QDW), i.e.\ a modulation of the quadrupolar ordering on length scales larger than the lattice period, analogues to the better known spin density waves (SDW) for dipolar spin states (e.g.~\cite{gruener94}) and charge density waves (CDW) for coupled electron density - lattice modulations (e.g.~\cite{gruener88}). These density waves are stationary ground states and hence to be distinguished from spin waves, which are collective moving excitations of spin degrees of freedom, whose transport properties are intensely studied with the development of spintronics~\cite{kajiwara10,zutic04,chumak15}. Nevertheless, a spatial displacement 
 of CDW and SDW demands zero energy and hence in principle they are free to move in any external perturbation, e.g.\ in an external field (sliding CDW or SDW, see e.g.~\cite{gruener88,gruener94,cox08} for reviews). This mechanism was originally proposed by Fr{\"o}hlich to explain superconductivity \cite{froehlich54}. However, in practice SDW and QDW are pinned by inhomogeneities of the material and a finite field is needed for depinning, leading also to excess noise,
e.g.~\cite{gruener88,gruener94,cox08}. More recently, the question of spontaneous time dependence and spontaneous motion was controversially discussed in the framework of time crystals and space-time crystals with proposals for perpetual motion in ion rings and structured ring-shaped BECs \cite{wilczek12,li12,robicheaux15} but no-go theorems seem to prevent this for the equilibrium ground states of a wide class of autonomous systems~\cite{nozieres13,bruno13,watanabe15}. The notion of dissipative time crystals for limit cycles in autonomous driven dissipative many-body systems was recently introduced in \cite{kessler21,kongkhambut22}.

Cold atom system have emerged as highly controllable simulators for aspects of magnetism and other condensed matter phenomena and time crystals (see e.g.~\cite{gross17,sacha18,guo20,daley22} for reviews). In this Letter, we use a diluted cloud of laser-cooled atoms submitted to optical feedback to generate light-mediated magnetic interactions. In this system, spontaneous spatial magnetic ordering occurs, with both dipole and quadrupole coupling~\cite{labeyrie18,kresic18,labeyrie22} depending on the magnitude and direction of an applied magnetic field.
We stress that these are not pseudo-spins in synthetic magnetic fields, but real magnetic moments in actual B-fields, however a strong coupling is provided by light-mediated interactions so that spontaneous magnetic ordering can emerge in a system which is neither very dense (condensed matter) or ultracold (quantum degenerate gases).
We report here the observation of a spontaneously sliding coupled SDW-QDW whose velocity is set by the magnetic field strength.


In the 1990's, several groups observed spontaneous pattern formation in hot atomic vapors using a retro-reflected laser beam~\cite{grynberg94, ackemann94}. For the experiments performed with an effective spin-1/2 structure, these observations corresponded to stationary magnetic dipole ordering due to Zeeman pumping~\cite{ackemann94, aumann97}. More recently, we have shown using cold $^{87}$Rb gases with a more complex energy level structure (corresponding to the $F = 2 \rightarrow F^\prime = 3$ transition of the D2 line) that quadrupole interaction terms can play a role in the observed spontaneous ordering~\cite{labeyrie18,labeyrie22}. These terms are associated with the $\Delta$m$_F$ = 2 Zeeman ground-state coherence induced by the laser fields, hence the name ''ground-state coherence'' (GSC) was given to this magnetic phase in which dipolar and quadrupolar degrees of freedom are excited.


\begin{figure}
\begin{center}
\includegraphics[width=1\columnwidth]{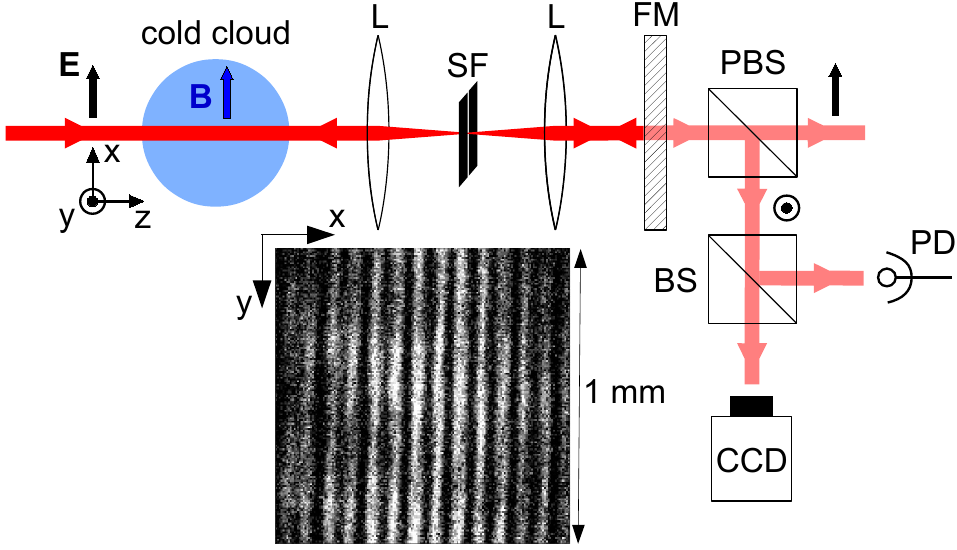}
\caption{Experimental scheme. A linearly-polarized, red-detuned laser beam is sent through a cloud of cold atoms and retro-reflected by a semi-transparent mirror (FM). An afocal telescope composed of two lenses (L) allows to generate an effective feedback mirror placed near the rear end of the cloud, inside the vacuum chamber (not shown). A weak magnetic field \textbf{B} is applied to the atoms along the direction of the input beam's polarization \textbf{E}. A slit (SF) placed inside the telescope allows to select the transverse wave vectors participating to the optical feedback. The light transmitted by the mirror passes through a polarizing beamsplitter (PBS) to select the polarization orthogonal to that of the input beam, and is split in two by a beamsplitter (BS). One part is sent to a photodiode (PD) to monitor its temporal fluctuations, while the other is sent onto a CCD. The typical spatial distribution of the light intensity corresponds to stripes, as shown in the inset.   Parameters: I$_0$ = 4.7 mW/cm$^2$, optical density in line center $b_0=130$, $\delta$ = -10$\Gamma$, B$_x = 0.14$ G, $\tau_{int} = 2$~$\mu$s. The stripe period is 77.6~$\mu$m$=\Lambda_c/2$.}
\label{fig:setup}
\end{center}
\end{figure}


The experimental setup, based on the single feedback mirror (SFM) scheme~\cite{firth90a,dalessandro91}, 
is sketched in Fig.~\ref{fig:setup} (see also e.g. Ref.~\cite{labeyrie18}). The nonlinear optical medium is a large, centimeter-sized cloud of cold $^{87}$Rb released from a magneto-optical trap. It is illuminated for 1 ms by a 1.8 mm-waist laser beam detuned to the red of the $F = 2 \rightarrow F^\prime = 3$ transition of the D2 line by $-10~\Gamma$, where $\Gamma$ is the linewidth of the transition. After traversing the cloud the laser beam is retro-reflected by a mirror, causing an optical feedback leading to the self-organization of both atomic susceptibility and light intensity in the transverse plane ($x$, $y$).  The spatial period $\Lambda_c$ of the emerging structure is determined by the diffractive dephasing between the on-axis pump and the off-axis spontaneously generated sidebands forming the spatial structure \cite{ackemann21}. The light intensity distribution in the cross-section of the beam is imaged on a camera placed behind the (semi-transparent) mirror. A photodiode gives access to the temporal behavior of the transmitted light.
All the signals presented here are detected in the linear polarization channel orthogonal to that of the incident light, termed ''lin $\perp$'' hereafter, as a signal in this channel indicates the emergence of the instability on zero background.

The GSC phase is usually composed of several domains with stripes oriented along different directions (see Fig.3 of Ref.~\cite{labeyrie22}, Figs.~12, 13 of Ref.~\cite{ackemann21}  and Ref.~\cite{labeyrie23s}).
We use a ``quasi-1D geometry'' (2D in real space, 1D in Fourier space) to facilitate the interpretation of the dynamic measurements. For this, we insert a spatial filter (SF) into the Fourier plane at the middle of an afocal telescope positioned between the cloud and the feedback mirror. In the SFM scheme, the wavenumber of the patterns is set by the distance between the medium and the feedback mirror due to diffraction, but all transverse wave vectors on a circle with this critical wavenumber can be excited. Using a simple slit as SF, two opposite wave vectors on this circle are selected. The resulting pattern thus corresponds to stripes oriented perpendicularly to the slit. Here, the stripes are oriented along $y$, as seen in Fig.~\ref{fig:setup}.

The numerical model used is based on a $F = 1 \rightarrow F^\prime = 2$ transition, simpler than the experimental $F = 2 \rightarrow F^\prime = 3$ to keep the number of coupled equations to solve reasonable, but complex enough to contain the necessary ingredients to explain the observed physics and in particular to allow for the existence of the $\Delta$m$_F$ = 2 ground-state coherence term, $\Phi=2\rho_{1 -1}$. These simulations give access to the time-resolved 2D distributions of atomic and light quantities in the transverse plane. The details of the model can be found in the supplementary material of Ref.~\cite{labeyrie18s,labeyrie23s}.

Fig.~\ref{fig:feedback_SDW} illustrates the dynamics of the spontaneous magnetic states. In the quasi-1D geometry explained before,  drifting stripes can be visualized  in the space-time diagrams in Fig.~\ref{fig:feedback_SDW}a-d) (see \cite{labeyrie23s} for animations). At a fixed spatial position, both dipole (Fig.~\ref{fig:feedback_SDW}d) and quadrupole (Fig.~\ref{fig:feedback_SDW}c)) are periodically oscillating in time. At a fixed time they form a modulated structure in space, consistent with a sliding multipole spin density waves.

For the dipole component, an anti-ferromagnetic state is connected to a periodic modulation of the longitudinal magnetization $w=\rho_{11}-\rho_{-1-1}=-m_z$ (Fig.~\ref{fig:feedback_SDW}d). It is given by the difference in occupation  $\rho_{11}$ and $\rho_{-1-1}$ in the Zeeman substates of the ground state with positive and negative magnetic quantum numbers.  Above a certain pump threshold, this structure emerge spontaneously from the unmodulated optical pump having linear input polarization, i.e.\ equal amounts of $\sigma_+$ and $\sigma_-$ light (and hence optical spin 0) everywhere and the thermal homogeneous and isotropic atomic cloud. It is sustained by a  spontaneously created optical spin structure (Fig.~\ref{fig:feedback_SDW}a shows the $\sigma_+$-component, the $\sigma_-$- lattice is complementary to this). In turn, this optical spin structure is sustained by spin selective scattering of the pump at the atomic orientation (see \cite{ackemann21,labeyrie23s} for details on the mechanism).
This anti-ferromagnetic state constitutes a SDW in the same way as anti-ferromagnetic ordering in systems with itinerant, i.e.\ delocalized, electrons \cite{slater51,overhauser61} can be due to SDW (see Fig.~S2 of \cite{labeyrie23s} for details).
However, this SDW is not stationary but drifts at constant speed. Figs.~\ref{fig:feedback_SDW}e, f) illustrates that the total magnetization $\vec{m}=(0,m_y,m_x$) has a screw-like behavior similar to circularly polarized light. This is due to the precession in the $x$-magnetic field (Fig.~\ref{fig:feedback_SDW}g). We will discuss details on frequencies, speeds and the spontaneous selection of drift direction and chirality below.

\begin{figure}
\begin{center}
\includegraphics[width=0.95\columnwidth]{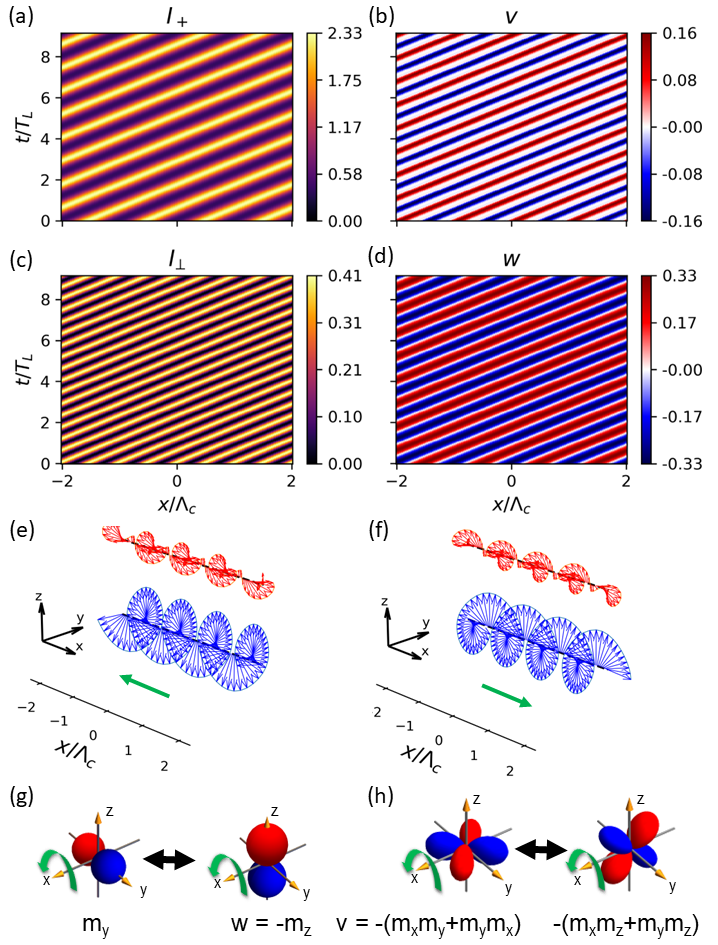}
\caption{a)-d) Space-time representation of multipolar spin density wave. Space is scaled in units of structure periods, time in units of the Larmor period. Parameters: $B_x=0.5$~G, $R=1$, OD$=70$, $\delta = -10 \Gamma_1$, $I_{\rm{in}} = 5$~mW/cm$^2$. 
a) Intensity of $\sigma_+$-component, b) Intensity of linear polarization component perpendicular to pump corresponding to experimental observable monitored, c) amplitude of quadrupolar state $v = -(m_x m_z+m_z m_x)$, d) amplitude of longitudinal orientation $w = -m_z$.  e,f) Illustration of precession of the magnetization $\vec{m}$ (lower, blue) and of  $(0,-(m_x m_y+m_y m_x),-(m_x m_z+m_z m_x))$ (upper red) representing one of the cones of the quadruple, for $B_x >0$. The wave in e) is drifting left in a right-handed screw, the one in f) right in a left-handed screw. Animations of the full wave dynamics are in \cite{labeyrie23s}.
g, h) Spherical harmonics illustrating the symmetry of the indicated state multipoles.
Red: positive (south pole), blue: negative values (north pole). The direction of precession is indicated by the green arrow.
}
\label{fig:feedback_SDW}
\end{center}
\end{figure}

Fig.~\ref{fig:feedback_SDW}b illustrates the structure in the quadrupolar component. $v = -2\Im{\rho_{-1 1}}$ is the imaginary part of the $\Delta m=2$-coherence between the stretched states and represents the quadrupole $m_x m_z+m_z m_x=-v$. It vanishes in the homogeneous state but gets excited around 0 in the structured state. The feedback mechanism is linked to the phase-sensitivity of the Raman coupling between the stretched states and hence an instability of the phase between the $\sigma_+$- and $\sigma_-$-components (see \cite{ackemann21,labeyrie23s}). Its dynamics is locked to the dynamics of the dipole components. Obviously, one cannot represent quadrupoles by a single vector but the trajectory of the tip of one of the quadrupolar cones (red in Figs.~\ref{fig:feedback_SDW}e, f) can be visualized as $(0,-(m_x m_y+m_y m_x),-(m_x m_z+m_z m_x))$  (see Fig.~\ref{fig:feedback_SDW}h for the precession of the quadrupole).



The ground states of SDW and CDW have a soft (Goldstone) mode structure and any small perturbation can thus bring this ground state into translational motion. In our experiment, this could be done in a controlled way by a minute tilt of the feedback mirror. However, again like SDW and CDW, these are usually pinned by experimental imperfections and a finite mirror tilt is needed to induce a drift \cite{seipenbusch97}.
 The striking numerical observation for the multipolar structures in the $J=1$-ground state is however that they \emph{drift spontaneously} even without a mirror tilt, i.e.\ they represent a spontaneously sliding multipolar spin density wave (SMSDW). We stress that this phenomenon does not occur for the other phases, in particular the anti-ferromagnetic phase existing around $B\approx 0$ and hexagonal and disordered phases that are observed, 
when the magnetic field is applied along the laser beam's propagation axis~\cite{kresic18,labeyrie18,kresic19}.


$w$ and $v$ as the primary relevant atomic variables oscillate at a frequency close to the Larmor frequency (about 15\% smaller, Fig.~\ref{fig:feedback_SDW}b,d, see also Fig.~\ref{fig:spectrum}). The $\sigma_\pm$-polarization components oscillate at the same frequency (Fig.~\ref{fig:feedback_SDW}a), whereas the linear polarization components orthogonal to the pump polarization used in the experimental detection scheme oscillates at twice this frequency.
\begin{figure}
\begin{center}
\includegraphics[width=0.95\columnwidth]{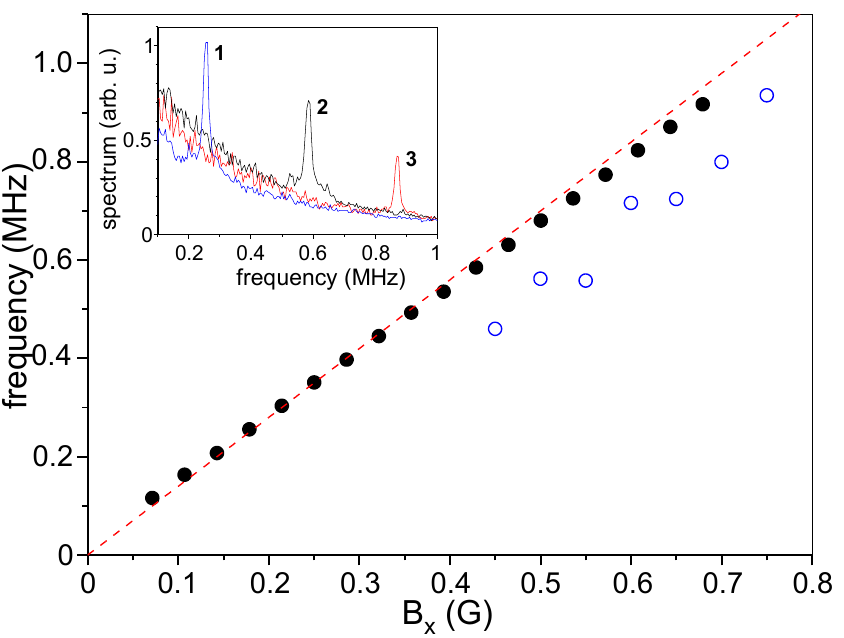}
\caption{\label{fig:spectrum}Temporal dynamics for  drifting structures in SMSDW phase. The frequency of the central frequency of the AC-peak is plotted as a function of $B_x$ for the parameters in Fig.~\ref{fig:setup} (black circles).  The frequency is obtained from spectra of the diffracted intensity (inset, \textbf{1}: $B_x = 0.25$ G, \textbf{2}: $B_x = 0.58$ G, \textbf{3}: $B_x = 0.87$ G).
Results from simulations are denoted by open blue circles with parameters OD$=130$, $\delta = -10 \Gamma_1$, $I_{\rm{in}} = 5$~mW/cm$^2$. 
The dotted red line indicates two times the Larmor frequency.
}
\label{fig:spectrum}
\end{center}
\end{figure}

For the verification of the predicted dynamics, we first looked at the temporal fluctuations of the diffracted light detected by the photodiode (see Fig.~\ref{fig:setup}). The Fourier transform of this signal shows a narrow peak whose position is proportional to B$_x$, as illustrated in Fig.~\ref{fig:spectrum}.  The frequencies observed are slightly lower than twice the Larmor frequency for high enough frequencies (twice, as the observation is done in the perpendicular polarization channel). This matches qualitatively the numerical observations although the observed reduction of frequency is about 15-27\% (compared to only 2-3\% in experiment), which merits further detailed characterization. The shift to lower frequencies is a quite common feature in dynamics with damping.

\begin{figure}
\begin{center}
\includegraphics[width=\columnwidth]{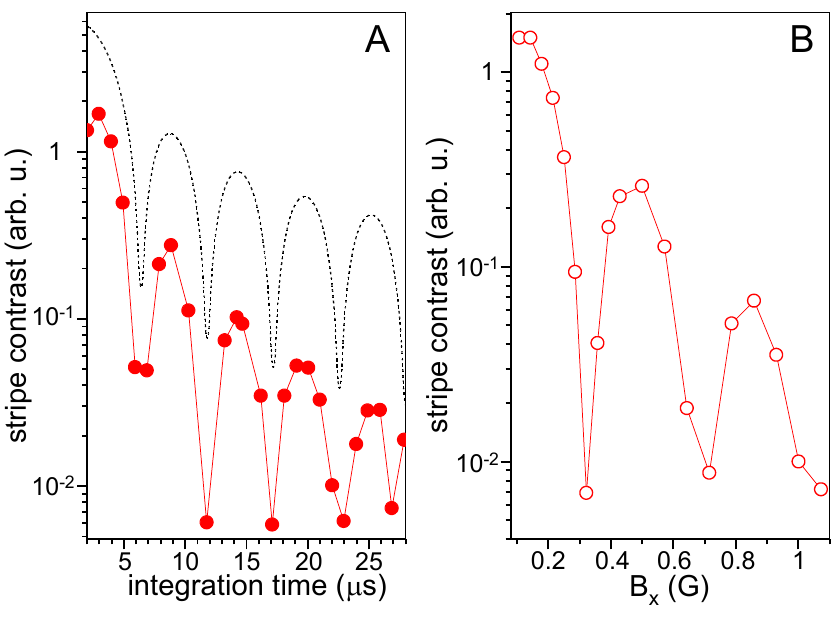}
\caption{SMSDW stripe drift. We plot the  stripe contrast (see text) versus: (A) camera integration time for $B_x = 0.14$ G; (B) magnetic field B$_x$ for $\tau_{int} = 2$~$\mu$s. Note the vertical logarithmic scale. The dots and circles (+ line to guide the eye) correspond to the experimental data. The dotted curve in (A) is the expectation for a drifting sinusoidal profile. }
\label{fig:drift}
\end{center}
\end{figure}

As camera equipment available to us does not allow a direct visualization of the drift, we recorded a series of forty images with an integration duration $\tau_{int}$, and studied how the contrast of the spatial modulation in these images varied with $\tau_{int}$. From this analysis, we inferred that the stripes do drift, and extracted the drift velocity. Due to the poor signal-to-noise ratio in the images and the fact that for each image the position (phase) of the stripes is different, we used the following procedure to quantify the contrast. We computed the Fourier transform of each image and then averaged the Fourier-transformed images. The amplitude of the peak corresponding to the wave vector of the stripes was taken as our measure of the contrast. Note that this quantity is proportional but not equal to the usual contrast used for interference fringes for instance, and can thus exceed one.

The result of this experiment is shown in Fig.~\ref{fig:drift}A, for B$_x = 0.14$ G. The dots correspond to the experimental data. The dotted curve shows the expected behavior for a drifting sine wave. The position of each minimum corresponds to a phase drift of $\Delta\varphi = n \times 2\pi$, with $n$ integer. It can be seen that the observed behaviors are qualitatively similar. For the experimental data, the time it takes for the wave to drift by one period is approximately 5.2 $\mu$s. This corresponds to a temporal modulation frequency of 0.19 MHz, consistent with the measurement presented in Fig.~\ref{fig:spectrum}. For a stripe period of $78$~$\mu$m, the corresponding drift velocity is 15~m/s.

The impact of the applied magnetic field B$_x$ on the drift velocity is illustrated in Fig.~\ref{fig:drift}B, where B$_x$ is varied from 0.1 to 1.1 G whilst the integration time is kept fixed at 2 $\mu$s. We observed again a pronounced variation of the contrast, with roughly equidistant minima. As discussed before, the spatial period of the SDW is set by the feedback mirror distance and does not vary with B$_x$. The observed behavior thus confirms the linear dependence of the drift velocity on B$_x$. Since the integration time is 2 $\mu$s, the modulation frequency corresponding to each minimum in Fig.~\ref{fig:drift}B is $n \times 500$ kHz, which is consistent with the data in Fig.~\ref{fig:spectrum}.

We now return to the discussion of the numerical observations.
As observed numerically and experimentally, the diffracted light detected in the orthogonal polarization channel is oscillating at two times the Larmor frequency (compare spacings of stripes in Figs.~\ref{fig:feedback_SDW}a and b), whereas the fundamental dynamics is at the Larmor frequency.
The appearance of the Larmor frequency as the oscillatory time scale is natural. It appears already in models for a $J=1/2$ ground state \cite{logvin98,huneus03} but was found to be damped for periodic patterns with homogeneous or nearly homogeneous pumping for experimentally accessible parameters. An additional observation from Fig.~\ref{fig:feedback_SDW}a, b is that $w$ and $v$ are anti-phased.
This is due to light induced anti-phased coupling terms between the dipole and quadrupole components \cite{labeyrie23s}. We conjecture that this additional oscillatory coupling absent in models with a $J=1/2$ ground state
provides the destabilization of the stationary periodic state to a time dependent one, as 
the simulations indicate that the Larmor oscillations become undamped if the Larmor frequency is of the order of or higher than this coupling frequency. (Note that dipolar and quadrupolar components are not coupled by the magnetic field directly but only via the light mediated coupling.) Interestingly, there is also a strong correlation between anti-ferromagnetic and quadrupolar ordering in the condensed-matter systems discussed, in particular in Ref.~\cite{onimaru05}.

The drift direction originates from spontaneous symmetry breaking. Starting simulations from noisy initial conditions, the direction of the drift (toward positive or negative $x$) is found with equal probability for both directions.
Interestingly, when we simulate of flip of the direction of B$_x$ during the run, we observe a systematic flip of the drift direction. This can be explained by the direction of precession.
Both the magnetization vector  $\vec{m}=(0,m_y,m_z)$ and $(0, -(m_x m_y + m_y m_x), -(m_x m_z + m_z m_x))$ form a left handed screw in space, if magnetic field and drift direction are parallel to each other, and a right handed screw,  if magnetic field and drift direction are anti-parallel to each other (Fig.~\ref{fig:feedback_SDW}e,f), details in \cite{labeyrie23s}).
This demonstrates chiral behaviour. If the direction of $B_x$ is suddenly flipped, the dynamic state can avoid a reshuffling of the sequence of states by switching the drift direction. This indicates that chirality is the decisive degree of freedom originating from spontaneous symmetry breaking, not drift direction.

Spontaneously drifting structures are also obtained if the Fourier filter is oriented along the $y$-axis, i.e.\ orthogonal to the applied magnetic field \cite{labeyrie23s}. Here the plane of precession is not orthogonal to the drift direction $x$ but in the plane spanned by the drift direction and the $z$-axis. Again, spontaneous symmetry breaking leads to chiral behaviour corresponding to the situation in the field of chiral quantum optics \cite{lodahl17} where the longitudinal polarization component of a strongly non-paraxial light fields and one transverse one couple to elliptically polarized light with the sense of rotation depending on propagation direction.

The observations constitute the demonstration of a novel spontaneous transport process in an out-of-equilibrium magnetic system. From a dynamical point of view, it extends the notion of spontaneous time dependencies in the dissipative time crystals discussed for the superradiant structures in transversely pumped cavities  \cite{kessler21,kongkhambut22} from oscillating to  drifting states. There is an interesting close phenomenological similarity between the cold atom system discussed here and the condensed matter systems discussed in \cite{onimaru05,dabrowski14,jang17,tazai19,hafner22}, including the strong link between anti-ferromagnetic ordering (SDW) and quadrupolar ordering (QDW). Even if a direct connection between the condensed-matter Hamiltonians and the cold atom system cannot be established at this stage, the analogy appears to be fruitful, in particular as most experiments in condensed-matter magnetism rely on inferring the structure from macroscopic observables like magnetic susceptibilities and not direct visualization, e.g. \cite{hafner22}. In condensed matter systems, SDW, QDW and the related CDG for non-magnetic systems are stationary ground states and drift only after strong enough parity breaking by external fields. The observation of spontaneous drift in a non-equilibrium version of magnetic ordering can be expected to trigger further fruitful research and insight in the question of time crystals and dissipative time crystals in general and spontaneous spin transport in particular.


\begin{acknowledgments}
The collaboration between the two groups is supported by the CNRS-funded Laboratoire International Associé (LIA) ''Solace'', and the Global Engagement Fund of the University of Strathclyde. We are grateful for useful discussions with Gian-Luca Oppo, Robert Cameron, Benjamin Hourahine, Ivor Kresic, Rina Tazai and Daniel Hafner.
\end{acknowledgments}

\appendix
\section{Supplementary material}

\section{Theoretical model}
The model is presented in the Supplementary Material of \cite{labeyrie18,labeyrie18s,kresic18} but reproduced here for convenience and adapted to the specific questions discussed here.

The dynamics of the ground state magnetization is described by optical Bloch equations for the reduced density matrix~$\rho$. A component of the density matrix for states with ground state magnetic quantum numbers $m=i,j$ is denoted by $\rho_{ij}$. Although the experiment is performed on a $F=2\to F'=3$ transition, the model is developed for a $F=1\to F'=2$ transition, which retains the properties of a $F \to F'=F+1$ transition as well as both dipole and quadrupole multipole components. Retaining the octupolar components (irreducible tensor rank 3) possible in the $F=2$ ground state would result in a system of twenty-four equations, i.e\ considerably more complex than the eight equations discussed below in (\ref{eqs:transfields}). We do not expect these additional components to be important for the dynamics although there might be quantitative corrections. First, they do not provide feedback to the light field  in the electric dipole approximation as the difference in tensor rank is 2 \cite{omont77}. Second, the magnetic field is not coupling manifolds with different rank $\kappa$, e.g \cite{stoeckmann14}, i.e.\ they will be only populated by secondary optical pumping processes.

 We take the quantization axis along the wavevector of the pump beam. Consequently the light fields are expressed in circular components by
\begin{equation}\label{eq:electricfield}
    \mathbf{E}(t)=\frac{1}{2}\sum_{q=\pm 1}(-1)^qE_q(t)\mathbf{\hat{e}_{-q}}e^{i\omega t}+\mbox{c.c.},
\end{equation}
where $\mathbf{\hat{e}_{\pm}}$ are the $\sigma^\pm$ polarization unit vectors.

Following the work of $\,$ \cite{mitschke86,dalibard89}, we make the following approximations:
\begin{itemize}
\item As the decay of the excited state populations and coherences is faster than the ones of the ground state, these are adiabatically eliminated, keeping terms to first order in $\Omega_\pm / \delta$, where $\Omega_\pm$ are the Rabi frequencies of the $\sigma^\pm$ fields, and $\delta$ is the laser beam detuning.
 \item Excited state populations are neglected as the pump rate is kept low. Hence the total population remains in the ground state and is constant, giving $\rho_{-1-1}+\rho_{00}+\rho_{11}=1$.
\item Optical coherences are adiabatically eliminated, keeping terms to first order in $\Omega_\pm / \delta$,.
\item We use the Land\'e $g$-factor $g_F=0.5$ of the $F=2$ ground state. The corresponding Larmor frequency $\Omega'_{x}$ is then given by
\begin{equation}
\Omega'_{x} = \Omega_{x} / \Gamma_2 = 0.23 \times B_{x}/{\mbox{G}}
\end{equation}
where $\Gamma_2$ is the coherence decay rate and half of the atomic linewidth $\Gamma$.
\item For simplicity in calculating the change in detuning, the Land\'e of the excited state is assumed also to be 0.5.
\item The total feedback length is less than 1 m and hence the retardation time for the feedback less than 3 ns. This is much shorter than the lifetime of the excited state of 27 ns and considerably shorter than the time scales of the magnetic dynamics of microseconds and above. Hence all retardation effects are neglected and the feedback to be assumed instantaneous.
\end{itemize}

The detuning and the Rabi frequencies are written in units of $\Gamma_2$, i.e. $\Delta=\delta/\Gamma_2$ and $\Omega'_\pm=\Omega_\pm / \Gamma_2$. The pump rates $P_\pm$ for the $\sigma_{\pm}$ fields coupling to stretched state transitions $m_1\to m_{2'}$ and $m_{-1}\to m_{-2'}$ are given by
\begin{equation}\label{eq:pumprates}
P_\pm= \frac{|\Omega'_\pm|^2} {1+\Delta^2} =
\frac{I_\pm}{I_{sat}}\,\frac{2}{1+\Delta^2}
\end{equation}
where $I_\pm$ are the intensities of the circularly polarized components and $I_{sat}$ is the saturation intensity. We consider $\Gamma_2=\pi\times 6.066$ MHz and $I_{sat}=1.669$ mW/cm$^2$ for circular light probing the $F=2\to F'=3$ transition of the D$_2$ line of $^{87}$Rb for all atoms in the stretched state $|m|=2$ (see \cite{steck10_87}). We also consider the sum and difference pump rates $ \mathcal{S}=P_++P_-,\, \mathcal{D}=P_+-P_- .$

The Raman transition pump rates $P_{\Lambda\pm}$ driving the $|\Delta m|=2$ coherence are given by
\begin{equation}
\label{eq:var_drivers}P_{\Lambda+}=\frac{2 \, \mbox{Re}({\Omega'}_+^{*}\Omega'_-)}{1+\Delta^2},\;
P_{\Lambda-}=- \frac{2 \, \mbox{Im}({\Omega'}_+^{*}\Omega'_-)}{1+\Delta^2}.
\end{equation}

Defining the system variables as
\begin{equation}
\begin{array}{l}
u=\rho_{-11}+\rho_{1-1}, \\
v=i(\rho_{-11}-\rho_{1-1}),\\
w=\rho_{11}-\rho_{-1-1},  \\
X=\rho_{11}+\rho_{-1-1}-2\rho_{00}, \\
y_1=\rho_{-10}+\rho_{0-1}, \\
z_1=\rho_{01}+\rho_{10},\\
y_2=i(\rho_{-10}-\rho_{0-1}), \\
z_2=i(\rho_{01}-\rho_{10}),\\
\label{atomvars}
\end{array}
\end{equation}
we derive a set of 8 coupled evolution equations (where ($\dot{\ }) \equiv \frac{d}{dt}(  )$):
\begin{equation}
\label{eqs:transfields}
\begin{array}{cll}
\dot{u} & = & -\Gamma_c u
+\left(\frac{5}{6}\mathcal{D} \Delta\right)v+\frac{1}{6}
P_{\Lambda-} \Delta w-\frac{1}{9}P_{\Lambda+}X+\frac{5}{18}P_{\Lambda+} \\
&& -\Omega'_x (z_2-y_2)/\sqrt{2} 
, \\
\dot{v} & = & -\Gamma_c v 
-\left(\frac{5}{6}\mathcal{D} \Delta \right)u
+\frac{1}{6}P_{\Lambda+} \Delta w+\frac{1}{9}P_{\Lambda_-}X-\frac{5}{18}P_{\Lambda-} \\
 & & +\Omega'_x (z_1-y_1) / \sqrt{2} 
,\\
\dot{w} & = & -\Gamma_w w-\frac{1}{6}P_{\Lambda-} \Delta u-\frac{1}{6}P_{\Lambda+} \Delta v-\frac{1}{9}\mathcal{D}X+\frac{5}{18}\mathcal{D} \\
 && -\Omega'_x (y_2+z_2) /\sqrt{2} 
,\\
\dot{X} & = & -\Gamma_X X-\frac{1}{3}P_{\Lambda+} u+\frac{1}{3}P_{\Lambda-} v+\frac{1}{3}\mathcal{D}w+\frac{5}{18}\mathcal{S} \\
 && +3 \Omega'_x (y_2-z_2) /\sqrt{2} 
,\\
\dot{y}_1 & = & -\Gamma_yy_1
+\Delta \mathcal{D}_y y_2
+\left(\frac{P_-'}{6}+\frac{1}{12}\left(\Delta P_{\Lambda-}-P_{\Lambda+}\right)\right)z_1\\
& & +\left(\frac{\Delta P_-'}{6}+\frac{1}{12}\left(\Delta P_{\Lambda+}+P_{\Lambda-}\right)\right)z_2 \\
&& +\Omega'_x v /\sqrt{2} 
,	\\
\dot{y}_2 & = & -\Gamma_yy_2
-\Delta \mathcal{D}_y y_1
-\left(\frac{\Delta P_-'}{6}-\frac{1}{12}\left(\Delta P_{\Lambda+}+P_{\Lambda-}\right)\right)z_1 \\ & & +\left(\frac{P_-'}{6}+\frac{1}{12}\left(P_{\Lambda+}-\Delta P_{\Lambda-}\right)\right)z_2 \\
 && +\Omega'_x (w-x-u) / \sqrt{2} 
,\\
\dot{z}_1 & = & -\Gamma_zz_1
+\Delta \mathcal{D}_z z_2
+\left(\frac{P_+'}{6}-\frac{1}{12}\left(\Delta P_{\Lambda-}+P_{\Lambda+}\right)\right)y_1 \\ & & -\left(\frac{\Delta P_+'}{6}+\frac{1}{12}\left(\Delta P_{\Lambda+}-P_{\Lambda-}\right)\right)y_2\\
&&-\Omega'_x v /\sqrt{2} 
,	\\
\dot{z}_2 & = & -\Gamma_zz_2
-\Delta \mathcal{D}_z z_1
+\left(\frac{\Delta P_+'}{6}-
\frac{1}{12} \left(\Delta P_{\Lambda+}-P_{\Lambda-}\right)\right)y_1 \\ & & +\left(\frac{P_+'}{6}+\frac{1}{12}
\left(\Delta P_{\Lambda-}+P_{\Lambda+}\right)\right)y_2 \\
 && +\Omega'_x (w+x+u)/\sqrt{2} 
  \, .
\end{array}
\end{equation}
The decay rates of the atomic variables are
\begin{eqnarray} \label{eq:relaxation}
\Gamma_w & = &r+\frac{1}{6}(P_++P_-),\\
\Gamma_X & = &r+\frac{11}{18}(P_++P_-),\\
\Gamma_c & = & r + \frac{7}{6}(P_++P_-) \\
& & -\frac{|\Omega'_+|^2+|\Omega'_-|^2}{3\left(1+\Delta^2 \right)},\\
\Gamma_y &= &r+P_+'+\frac{7}{12}P_-' \quad\\
 \Gamma_z & = &r +\frac{7}{12}P_+'+P_-',\,\mbox{with}\\
  P_{\pm}'&=&\frac{|\Omega'_\pm|^2} {1+\Delta^2} \, ,
\end{eqnarray}
where $r$ is an effective decay rate of the Zeeman ground state population and coherences. Its lower limit results from the residual atomic motion leading to a wash-out of the structures and can be estimated to be about $2.8\times 10^3$s$^{-1}$, i.e.\ $r\ \approx 1.5\times 10^{-4}$ in the scaled units used here. The difference pump rates in the light-shift terms for $y_1$, $y_2$, $z_1$, $z_2$ are
\begin{equation}
\begin{array}{cc}
    \mathcal{D}_y=P_+'-\frac{7}{12}P_-', &\mathcal{D}_z=\frac{7}{12}P_+'-P_-'.
\end{array}
\end{equation}

To summarize, the orientation $w$ is driven by the difference pump rate $D$ leading to optical pumping, the longitudinal alignment by the total pump rate $S$ leading to a population of the stretched states overall and the real and imaginary part of the transverse alignment $\Phi$ via the phase dependent pump rates $P_{\Lambda\_\pm}$. This is the same phase sensitivity as in processes related to electromagnetically induced transparency \cite{fleischhauer05}.

The equations for the evolution of the amplitudes $E_\pm$ of the forward beam through the diffractively thin cloud are
\begin{equation}\label{eq:evol}
\begin{array}{l}
\frac{\partial} {\partial z} E_\pm=i \chi_\pm \frac {k}{2} \left[ \left(1\pm\frac{3}{4}w+\frac{1}{20}X\right)E_\pm+
\frac{3}{20}(u\mp iv)E_\mp\right], \\
\end{array}
\end{equation}
where the linear susceptibility $\chi_\pm$ is
\begin{equation} \label{eq:chi}
\chi_\pm=\frac{OD}{k L}\frac{i+\Delta \mp \Omega'_z}{1+\Delta^2},
\end{equation}
where $OD$ is the optical density, since in simulations we include both light absorption and refraction, and the linear and non-linear Faraday effects. Formulas (\ref{eq:evol}) and (\ref{eq:chi}) are used in the main paper with un-normalized variables.

After traversing the cloud, the beams propagate a distance of two times the mirror distance $d$ (to the feedback mirror and back), which is governed by
\begin{equation} \label{eq:prop}
    \frac{\partial}{\partial z} E_\pm=-\frac{i}{2k} \Delta_\perp E_\pm,
\end{equation}
where $\Delta_\perp$ is the partial derivative in the transverse coordinates. We take the interacting Rabi frequencies as the sum of the forward field at the entrance of the medium and the reentrant field at the exit field of the medium calculated from Eqs.~(\ref{eq:evol}), (\ref{eq:prop}) neglecting the wavelength-scale grating resulting from the interference of counterpropagating fields. Both assumptions proved to be suitable in earlier studies in cold atoms \cite{labeyrie14}. In particular, as the dynamics is evolving on time scales of the order of $1/r$ and the period of the wavelength scale grating is about a factor of 100 smaller than the pattern period, atomic motion is expected to provide a strong damping to the wavelength-scale modulations (Supplementary material of \cite{labeyrie14}). The details of numerical procedures used in our simulations are given in Ref. \cite{tesio14t}.

Equation~(\ref{eq:prop}) is solved in Fourier space, i.e. the reentrant field after propagating a distance $2d$ in Fourier space, $\tilde{E}(q_x,q_y,2d)$, is given by
\begin{equation} \label{eq:fourier}
\tilde{E}_\pm (q_x,q_y,2d) = e^{i\frac{q^2 2d}{2k}}\,  \tilde{E}_\pm (q_x,q_y,0)
\end{equation}
with $q=\sqrt{q_x^2+q_y^2}$. Minimum threshold is reached for the wavenumber $q_c=2\pi/\Lambda_c$ for which the diffractive phasor is $i$, as this provides optimal conversion of the phase modulation imprinted onto the transmitted field by the structure in the atoms into an amplitude modulation, see Sec.~`Mechanism of instability and connection to spin density waves'. In the experiment discussed here $\Lambda_c \approx 155\, \mu$m (in the perpendicular polarization to the pump $\Lambda_c/2 \approx 76\, \mu$m is measured, see Figs.~2a, c of the main text and Fig.~S2a below for the difference in length scale). This corresponds to a mirror distance $d\approx - 15$~mm. 

\section{Irreducible tensor components}\label{sec:irreducible}
A useful basis to obtain insight in insight into light matter interactions and spin precession are often the irreducible tensor operators $\mathcal{T}_q^\kappa$ with rank $\kappa$ \cite{omont77,rochester10t,stoeckmann14}. The operators have rotational symmetries of real and imaginary parts of spherical harmonic functions with angular momentum numbers $l=\kappa$ and projection numbers $m=q$. For $F=1$ Zeeman ground system, the terms are: rank-0 representing the monopole or total population, rank-1 representing the dipole components or orientations and rank-2 representing the quadrupole components or alignments. In the framework of the dynamical variables used in (\ref{atomvars}), these are
\begin{equation}
\begin{array}{l}
\rho_{0}^0=\rho_{-1 -1}+ \rho_{0 0} + \rho_{1 1} =1 \\
\rho_{-1}^1=-\frac{1}{2\sqrt{2}}(y_1+z_1+i(y_2+z_2)) \\
\rho_0^1=\frac{1}{\sqrt{2}}w,\\
\rho_{+1}^1=\frac{1}{2\sqrt{2}}(y_1+z_1-i(y_2+z_2))\\
 \rho_{-2}^2=\frac{1}{2}(u+iv) \\
 \rho_{-1}^2=\frac{1}{2\sqrt{2}}(z_1-y_1+i(z_2-y_2)) \\
\rho_0^2=\frac{1}{\sqrt{6}}X,\\
 \rho_{+1}^2=\frac{1}{2\sqrt{2}}(y_1-z_1+i(z_2-y_2)) \\
 \rho_{+2}^2=\frac{1}{2}(u-iv) \\
\label{tenscomps1_used}
\end{array}
\end{equation}

\begin{figure}
\begin{center}
\includegraphics[width=\columnwidth]{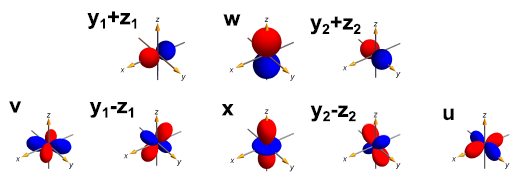}
\caption{\label{fig:magmoms} Spherical harmonics determining the symmetry of irreducible tensor components related to the indicated state multipoles. Red: positive, blue: negative values. From \cite{kresic17t}. \emph{Courtesy of Ivor Kresic.}}
\end{center}
\end{figure}
In the $\kappa =1$-manifold, one can define the Cartesian components of the magnetization $\vec{m}=(m_x,m_y,m_z)$ (see Fig.~\ref{fig:magmoms}) as
\begin{equation}
\begin{array}{l}
m_x : = -\,\frac{1}{\sqrt{2}}\, \left(y_1+z_1\right)\\
m_y : = \frac{1}{\sqrt{2}}\, \left(y_2+z_2\right)\\
m_z : = -\, w\\
\label{eq:tensor1}
\end{array}
\end{equation}
leading to the well known precession equations in a $B_x$-field
\begin{equation}
\begin{array}{l}
\dot{m}_y= -\Omega_x' m_z\\
\dot{m}_z=\Omega_x' m_y
\label{eq:precession1}
\end{array}
\end{equation}
or
\begin{equation}
\dot{\vec{m}} = (\Omega_x',0,0) \times \vec{m}
\label{eq:precession1vector}
\end{equation}
with the solution
\begin{equation}
\begin{array}{l}
m_y= m_0 \cos{(\Omega_x't\pm \frac{\pi}{2})}\\
m_z = m_0 \cos{(\Omega_x't)} .
\label{eq:precession1_solution}
\end{array}
\end{equation}
Here $m_0$ is the amplitude of the magnetization and the phase depends on whether $B_x$ is positive or negative.

In the $\kappa=2$-manifold, the tensor components are linked to quadrupolar magnetization components (\cite{rochester10t,stoeckmann14}, see Fig.~\ref{fig:magmoms}) as
\begin{equation}
\begin{array}{l}
y_1-z_1 = -\sqrt{2}\,(m_x m_z + m_z m_x)\\
y_2-z_2 = \sqrt{2}\, (m_y m_z + m_z m_y)\\
u  =  m_x^2 -m_y^2 \\
v  = -(m_x m_y + m_y m_x) \\
X  =   3m_z^2 -m^2.\\
\label{eq:tensor2}
\end{array}
\end{equation}
and the  precession equations are
\begin{equation}
\begin{array}{lll}
\dot{y}_1-\dot{z}_1 & = & \sqrt{2}\Omega_x' v\\
\dot{y}_2-\dot{z}_2 & = & -\sqrt{2}\Omega_x' X -\sqrt{2} \Omega_x' u \\
\dot{u} & = & \frac{1}{\sqrt{2}}\Omega_x' (y_2-z_2) \\
\dot{v} & = & -\frac{1}{\sqrt{2}}\Omega_x' (y_1-z_1) \\
\dot{X} & = & \frac{3}{\sqrt{2}}\Omega_x' (y_2-z_2)  ,\\
\label{eq:precession2}
\end{array}
\end{equation}
Solutions are
\begin{equation}
\begin{array}{lll}
 v & = -(m_x m_y + m_y m_x) & = v_0\, \cos{(\Omega_x't)}\\
 \frac{y_1-z_1}{\sqrt{2}} &  = -\,(m_x m_z + m_z m_x) &  = v_0\, \cos{(\Omega_x't\mp \frac{\pi}{2})},\\
\label{eq:precession2_solution_1}
\end{array}
\end{equation}
and
\begin{equation}
\begin{array}{lll}
\frac{y_2-z_2}{\sqrt{2}} & = (m_y m_z + m_z m_y) & = q_0 \,\cos{(2\Omega_x't)})\\
u & = m_x^2 -m_y^2 & = \frac{1}{2}\,q_0 \,\cos{(2\Omega_x't\mp \frac{\pi}{2})}\\
 X & =3m_z^2 -m^2 & = \frac{3}{2}\,q_0\, \cos{(2\Omega_x't\mp \frac{\pi}{2})},\\
\label{eq:precession2_solution_2}
\end{array}
\end{equation}
where $v_0$, $q_0$ are the amplitudes of $v$, $(y_2-z_2)/\sqrt{2}$, respectively and the phases depend again on whether $B_x$ is positive or negative. The reason that $w$, $y_2+z_2$, $v$, $y_1-z_1$ oscillate at the Larmor frequency in a $B_x$-field, whereas $u$, $y_2-z_2$ and $X$ oscillate at twice the Larmor frequency is that the former need to rotate by 360$^\circ$  to return  to a state equivalent to their original state, whereas the latter need only to rotate by 180$^\circ$. This is due to their symmetry properties with respect to the $x$-axis (see Fig.~\ref{fig:magmoms}).

\section{Dynamics of $\Delta m=2$-coherence}\label{sec:coherence}

The complete solution of the system (\ref{eqs:transfields}) is out of the scope of this paper, but important insight on the role of the coherences and the associated magnetic quadrupoles can be drawn already from an inspection of the equations of motion for the coherence $\Phi=u+iv$ alone, neglecting the coupling to other moments and the light shift terms.

Taking the field as a superposition of $\Omega_+$ and $\Omega_- \exp{(-\phi_L)}$ with $\Omega_+$, $\Omega_-$ real and $\phi_L$ denoting the phase difference, one obtains for the Raman pump rates
\begin{eqnarray}
P_{\Lambda+} & = & \frac{2\Omega'_+ \Omega'_-}{1+\Delta^2} \,  \cos{\phi_L}\\
P_{\Lambda-} & = & \frac{2 \Omega'_+ \Omega'_-}{1+\Delta^2} \, \sin{\phi_L},
\end{eqnarray}
and for the stationary solutions
\begin{eqnarray}
  u &=& \frac{5}{18}\, \frac{1}{\Gamma_c}\,\frac{2\Omega'_+ \Omega'_-}{1+\Delta^2} \,  \cos{\phi_L} = \frac{5}{18}\, \frac{1}{\Gamma_c}\,P_{\Lambda+}
  \\
  v &=& - \frac{5}{18}\, \frac{1}{\Gamma_c}\, \frac{2\Omega'_+ \Omega'_-}{1+\Delta^2}\,  \sin{\phi_L} = - \frac{5}{18}\, \frac{1}{\Gamma_c}\,P_{\Lambda-}\\
  \Phi & = & \frac{5}{18}\, \frac{1}{\Gamma_c}\,\frac{2\Omega'_+ \Omega'_-}{1+\Delta^2} \, e^{-i\phi_L}\,  .
\end{eqnarray}
This shows that the phase $\phi_L$ between the circular polarization components determines the phase of the coherence $\Phi$. The angle of the polarization direction with respect to the $x$-axis, $\phi_p$, for linearly polarized light (or the principal axis for $\Omega_+\neq \Omega_-$) is related to half the phase difference as the phase $\phi_L$ varies between $0$ and $2\pi$ but the polarization direction $\phi_p$ only between $0$ and $\pi$:
\begin{equation}\label{eq:angle}
\phi_p= \, \frac{\phi_L-\pi}{2}.
\end{equation}
 The direction of the principal axis of the quadrupole $\phi_Q$ is linked to the phase of the coherence via an equation like (\ref{eq:angle}). Hence  the polarization direction of the light is directly controlling the direction of the quadrupole, $\phi_P=\phi_Q$, which is of course expected from symmetry arguments. For example, in the notation used the $x$-polarized input beam corresponds to $\phi_L=\pi$ and $\phi_p=0$.
It pumps (for $\Omega'_z=0$), $u\neq0$, $v=0$, i.e.\ the cones of the resulting quadrupole (right hand side, lower row of Fig.~\ref{fig:magmoms}) are directed along the $x$-axis, the orthogonal cones along $y$.  $\phi_L=\pi/2$ corresponds to light polarized at -45$^\circ$ to the $x$-axis. It pumps (for $\Omega'_z=0$), $u=0$, $v\neq 0$, i.e.\ the cones of the resulting quadrupole (left hand side, lower row of Fig.~\ref{fig:magmoms}) are directed at $45^\circ$  along the bisections of the  $x$-axis and $y$-axis. In between, there is a smooth transition stemming from the form of the spherical harmonic function. A situation with $\phi_L=\pi/4$ corresponding to a polarization direction of -22.5$^\circ$ gives $u=v$. Hence, the modulation of $v$ implies a modulation of the principal axis of the quadrupole.

The link between polarization direction, phase $\phi_L$ between the $\sigma_\pm$-polarization components and phase of the coherence $\Phi$ provides an additional and different feedback mechanism from the one leading to creation of the orientation $w$ via the amplitude differences between the $\sigma_\pm$-polarization components. A fluctuation in the light phase $\phi_L$ will perturb the phase of $\Phi$ and hence create some fluctuation in $v$. This couples via (\ref{eq:evol}) and the diffractive dephasing (\ref{eq:prop}) back onto the light phase (or a fluctuation in $\Phi$ creating some amount of $v$ feeds back into the light phase and via this into $\Phi$) \cite{ackemann21}.

\section{Mechanism of instability and connection to spin density waves}\label{sec:mechanism}

The numerical model used is based on a $F = 1 \rightarrow F^\prime = 2$ transition, simpler than the experimental $F = 2 \rightarrow F^\prime = 3$ to keep the number of coupled equations to solve reasonable, but complex enough to contain the necessary ingredients to explain the observed physics and in particular to allow for the existence of the $\Delta$m$_F$ = 2 ground-state coherence term, $\Phi=\rho_{1 -1}/2$. These simulations give access to the time-resolved 2D distributions of atomic and light quantities in the transverse plane.  Fig.~\ref{fig:feedback_SDW} illustrates the dynamics and stabilization mechanism of spontaneous magnetic states. An anti-ferromagnetic state is connected to a periodic modulation of the longitudinal magnetization $w=\rho_{11}-\rho_{-1-1}=-m_z$. It is given by the difference in occupation  $\rho_{11}$ and $\rho_{-1-1}$ in the Zeeman substates of the ground state with positive and negative magnetic quantum numbers (black line in Fig.~\ref{fig:feedback_SDW}b, it illustrates the space-profile in a snapshot, but due to the wave-like nature it can be also taken as an indicator of the temporal oscillation.)
$\sigma_+$-light has a stronger light-matter coupling in regions with $w>0$ and $\sigma_-$-light otherwise. Hence both polarization components will be phase modulated after traversing the cloud but these modulations will be anti-phased, i.e.\ the opposite of each other. The phase modulation is converted to amplitude modulation by diffraction in the feedback loop. In the reentrant light, both $\sigma_+$ and $\sigma_-$ light are spatially modulated but in anti-phase (red and blue lines in Fig.~\ref{fig:feedback_SDW}a). This closes the feedback loop (see, e.g., \cite{ackemann21} for details). An optical spin structure sustains an atomic spin structure and vice versa. Above a certain pump threshold, these structures emerge spontaneously from the unmodulated optical pump having linear input polarization, i.e.\ equal amounts of $\sigma_+$ and $\sigma_-$ light (and hence optical spin 0) everywhere and the homogeneous and isotropic atomic cloud.

 Based on the observation that anti-ferromagnetic ordering can be due to SDW in systems with itinerant, i.e.\ delocalized, electrons \cite{slater51,overhauser61}, an
interpretation in terms of SDW is possible as the profiles of $\rho_{11}$ and $\rho_{-1-1}$ would represent the corresponding modulated carrier densities with spin-up and spin-down (red and blue lines in Fig.~\ref{fig:feedback_SDW}b, we will discuss the symmetry breaking related to the second harmonic later). Recalling that a SDW can be also thought of as spatially anti-phased CDW for the respective carriers \cite{gruener94}, one can construct a further analogy. In the CDW, the carrier density modulation is stabilized by a corresponding modulation of the lattice atoms and vice versa. Here the role of the  carrier density is played by  the population of the Zeeman states $\rho_{\pm 11}$ and the role of the lattice modulation by the modulated optical intensities. Note that in first order (without harmonics) the total number of atoms, $\rho_{1 1}+\rho_{-1 -1}$ in the magnetic (also called stretched) states (green line in Fig.~\ref{fig:feedback_SDW}b) and the total optical intensity are not modulated, as expected for a SDW. The residual modulation at half the fundamental period is due to the fact that the total population in the stretched states peaks each time the system has maximum orientation $|w|$ and that during the inversion of the population also the population $\rho_{00}$ of the intermediate state changes (see black line for longitudinal alignment $X$ in Fig.~\ref{fig:feedback_SDW}d).

\begin{figure}
\begin{center}
\includegraphics[width=\columnwidth]{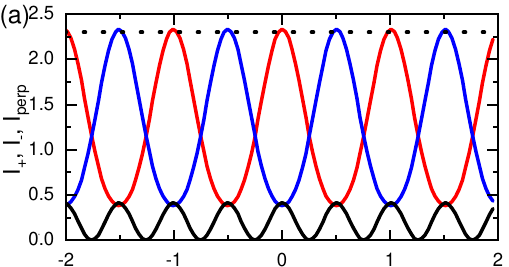}
\includegraphics[width=\columnwidth]{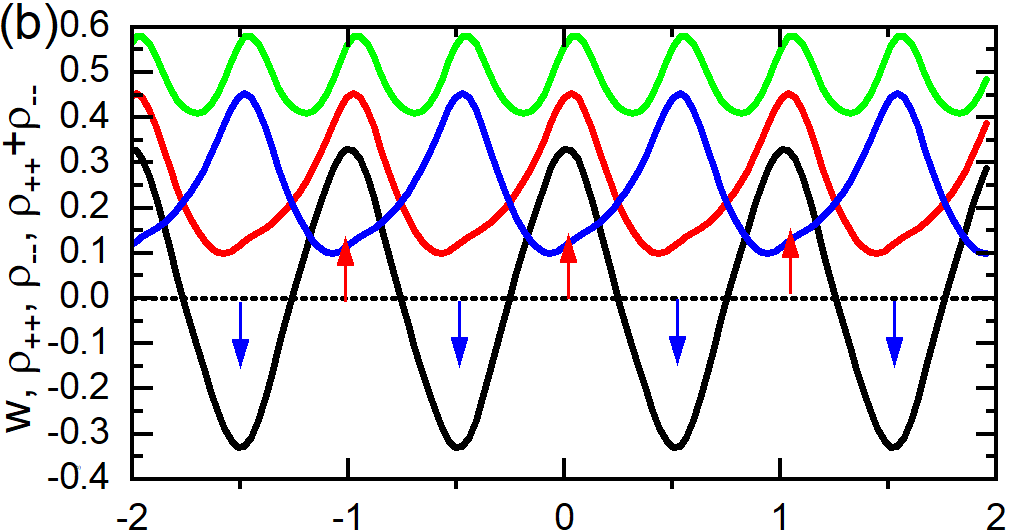}
\includegraphics[width=\columnwidth]{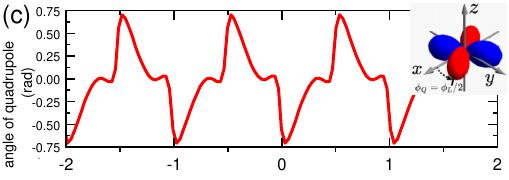}
\includegraphics[width=\columnwidth]{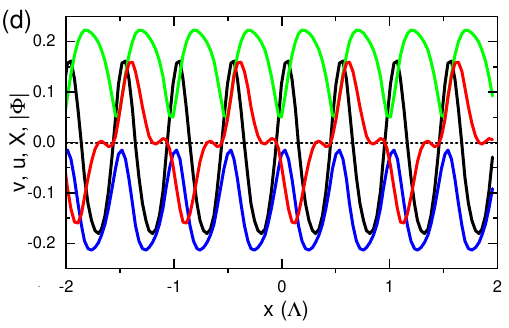}
\caption{Illustration of drifting multipole spin wave and mechanism of stabilization of anti-ferromagnetic state via light-mediated interactions and connection to SDW and CDG. Depicted are spatial profiles of snapshots of magnetic states and light intensities with space measured in units of the lattice period. Due to the wave-like nature, it can be also taken as an indication of the temporal structure. The wave is propagating to the right (positive $x$-direction).
a) Intensity profiles. Red: $\sigma_+$, blue: $\sigma_-$, black solid: perpendicular linear, black dotted: linear parallel to pump. b) Dipolar spin states. Red line: population of spin-up state $\rho_{11}$, blue: population of spin-down state $\rho_{-1-1}$, black: orientation $w=\rho_{1 1}-\rho_{-1 -1}=-m_z$, green: total density of atoms in stretched states $\rho_{11}+\rho_{-1-1}$.
c) Direction of quadrupole as calculated from Eq.~(\ref{eq:angle}). The angle of the red handle of the quadrupole to the $x$-axis is depicted. See text.
d) Quadrupolar states. Red: $v$, blue: $u$, black: $X$, green: $|\phi|$.  Parameters: $B_x=0.5$~G, $R=1$, OD$=70$, $\delta = -10 \Gamma$, $I_{\rm{in}}=5$ mW/cm$^2$.
}
\label{fig:feedback_SDW}
\end{center}
\end{figure}

Fig.~\ref{fig:feedback_SDW}d illustrates the structures in the quadrupolar components, $v = -(m_x m_z+m_z m_x)$ vanishes in the homogeneous state but gets excited symmetrically around 0 in the structured state (red line in Fig.~\ref{fig:feedback_SDW}d). The feedback mechanism is linked to the phase-sensitivity of the Raman coupling between the stretched states and hence an instability of the phase between the $\sigma_+$- and $\sigma_-$-components (see \cite{ackemann21,labeyrie18s}).
The optical intensity projected on the linear polarization state orthogonal to the pump is modulated at half the fundamental period (black line in Fig.~\ref{fig:feedback_SDW}a) The component remaining in the input polarization is essentially constant (dotted black line in Fig.~\ref{fig:feedback_SDW}a). As the latter pumps the components $u = m_x^2-m_y2$ and $X = 3 m_z^2-m^2$, these are already excited in the homogeneous states and hence acquire a modulation on top of an offset and at half the fundamental period as a secondary effect. (The average excitation of $X$ depends on parameters. Here the inversion symmetry is only slightly broken.)

As the result of the modulation of $u$ and $v$, the amplitude of the total coherence $|\Phi|$ is also modulated at two times the fundamental period and around an offset (green line in Fig.~\ref{fig:feedback_SDW}d). However, the main effect is that the direction of the quadrupole is modulated in space (and time), see Fig.~\ref{fig:feedback_SDW}c). Where $v$ is zero and only the $u$ pumped already in the homogeneous state is present, the orientation is along the $x$-axis for one pole and the $y$-axis for the other. In presence of $v$ it is rotated out of these directions.
Maximum deviation from $x$-axis is about 40$^\circ$, where $u$ is nearly zero and hence the quadrupole is dominated by the emerging $v$.

\section{Details on numerical results}

\begin{figure}
\begin{center}
\includegraphics[width=\columnwidth,clip=true]{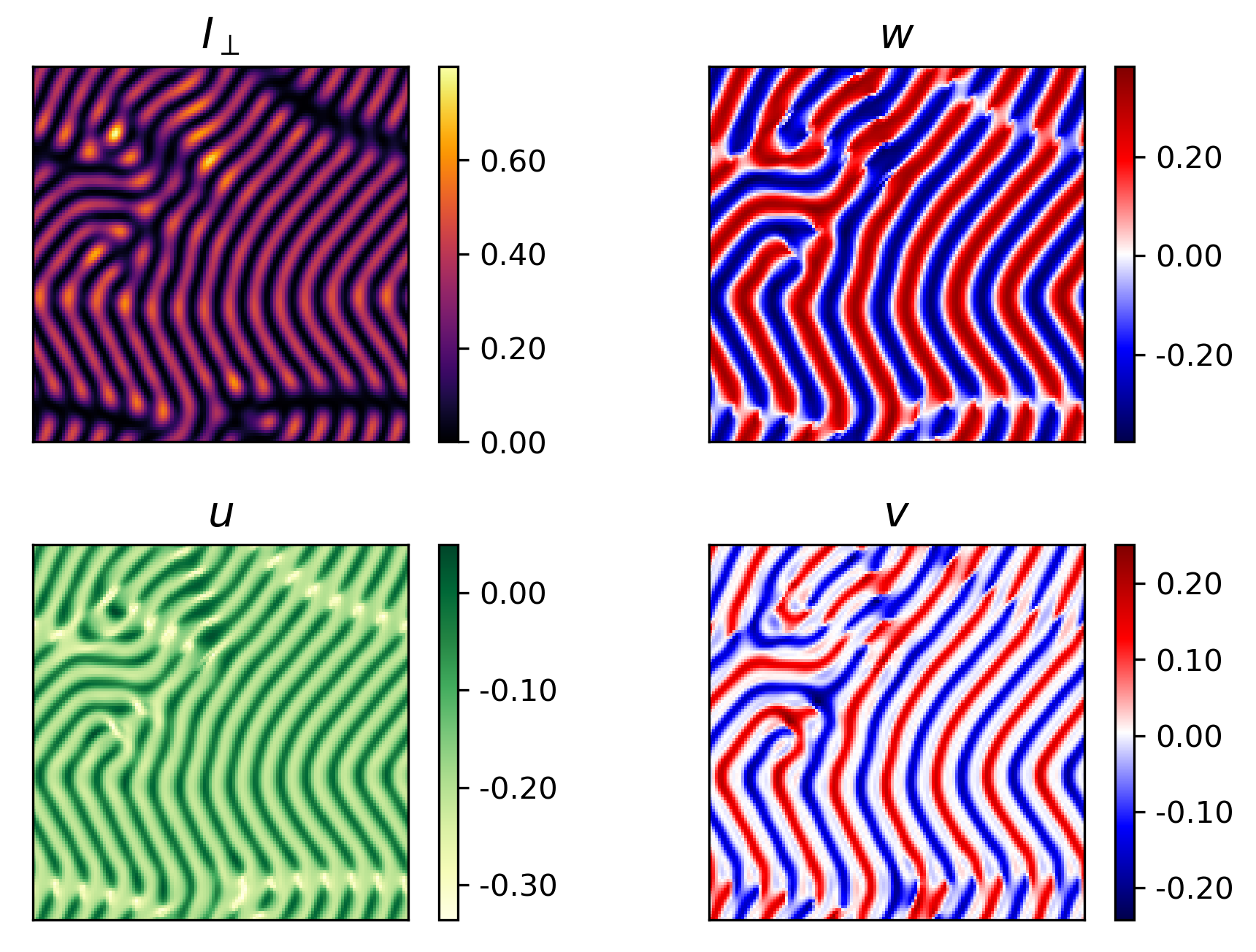}
\caption{Numerical example for structures obtained without Fourier filtering. a) Intensity in perpendicular component, b) $w$, c) $u$, d) $v$. Note that $w$, $v$ are modulated at the fundamental critical period and $I_\perp$ and $u$ at the harmonic. Parameters: $b_0$=70, pump intensity $I_{\rm{in}}=5$ mW/cm$^2$, $B_x=0.5$ G and detuning $\delta=-10\Gamma$. }
\label{fig:disorder}
\end{center}
\end{figure}

Unless the transverse extension of the pumped area is very small, i.e.\ only a few time the critical period of the structure, the resulting states is very disordered (Fig.~\ref{fig:disorder}) and consists of small patches with stripe-like structures with defects in between. Within the small patches the stripes are drifting (see animation \cite{labeyrie23s}, GSC-disordered.mp4 in 'movies`). Time averaged pictures like the ones experimentally taken are hence highly disordered and display in tendency the defect structures and potentially some remnants of modulation surviving time integration as integration time and oscillation period are not the same  (e.g., Fig. 3 of \cite{labeyrie18}, Fig. 12 of \cite{ackemann21}).

\begin{figure}
\begin{center}
\includegraphics[width=\columnwidth,clip=true]{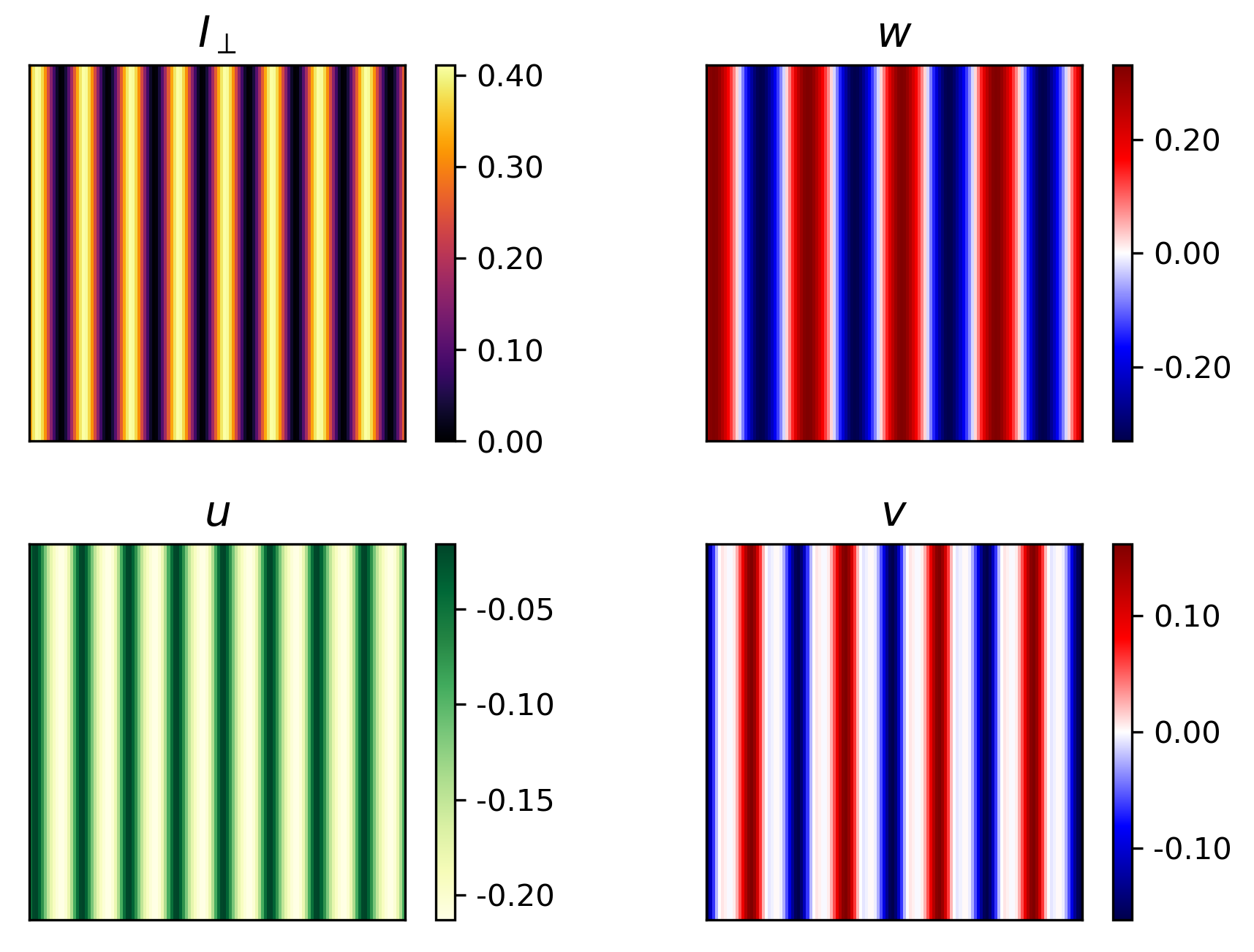}
\caption{Numerical example for structures obtained with Fourier filtering. The slit is oriented along the $x$-direction (horizontal), parallel to the $B$-field. a) Intensity in perpendicular component, b) $w$, c) $u$, d) $v$. Note that $w$, $v$ are modulated at the fundamental critical period and $I_\perp$ and $u$ at the harmonic. Parameters: $b_0$=70, pump intensity $I_{\rm{in}}=5$ mW/cm$^2$, $B_x=0.20$ G and detuning $\delta=-10\Gamma$. }
\label{fig:stripes}
\end{center}
\end{figure}

Implementing the Fourier filter, stripes are enforced (Fig.~\ref{fig:stripes}). These are drifting orthogonal to the stripe lines, parallel to the wavevector. For a slit in $x$-direction and hence vertical stripes, the drift is visualized in the animation in \cite{labeyrie23s} (vertical-rolls.mp4 in 'movies`). Sampling the perpendicular polarization component will yield double the frequency than when sampling a circular polarization component. The anti-phased behavior of $v$ and $w$ is also very clear.

\begin{figure}
\begin{center}
\includegraphics[width=\columnwidth,clip=true]{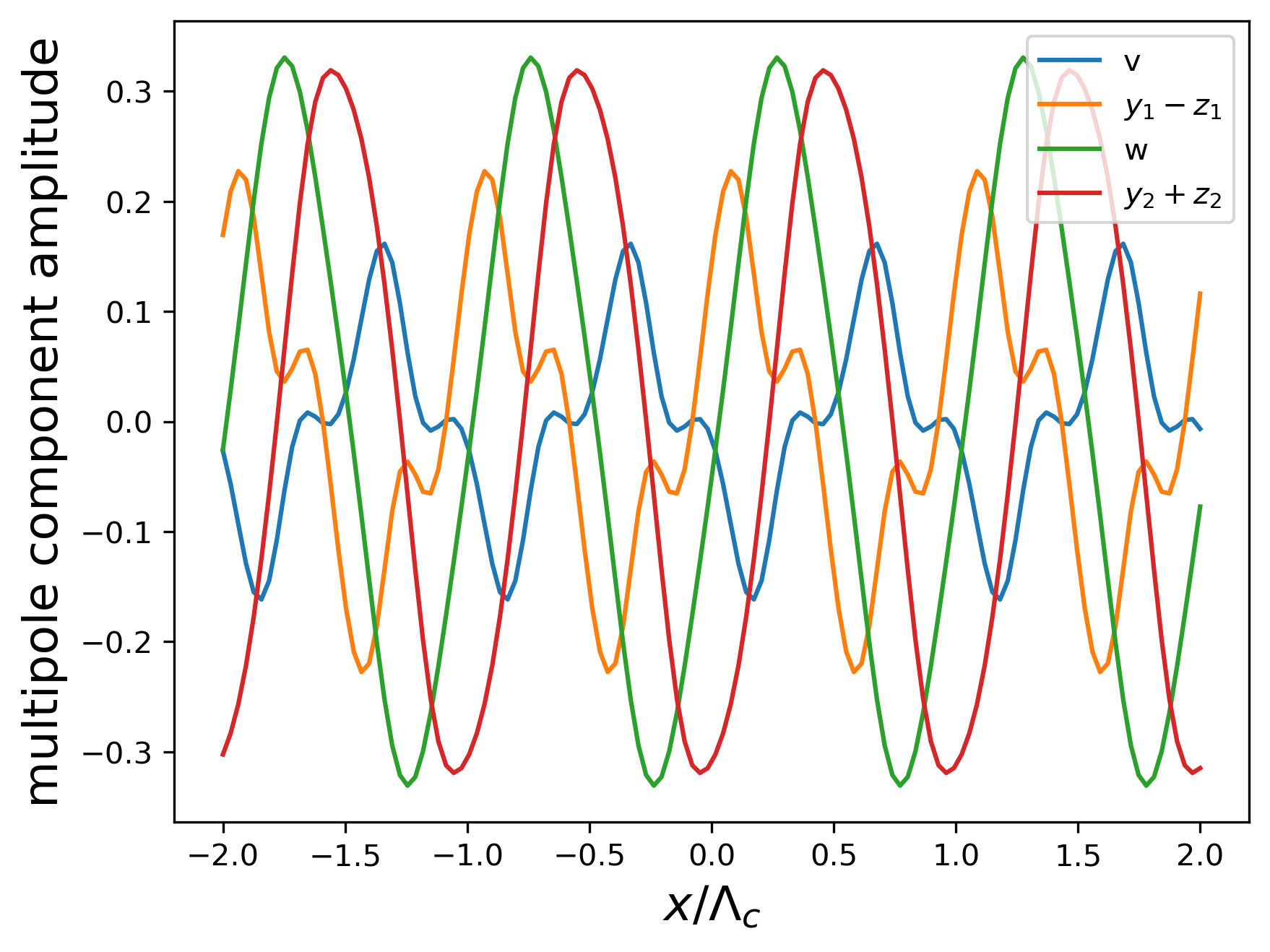}\\
\includegraphics[width=\columnwidth,clip=true]{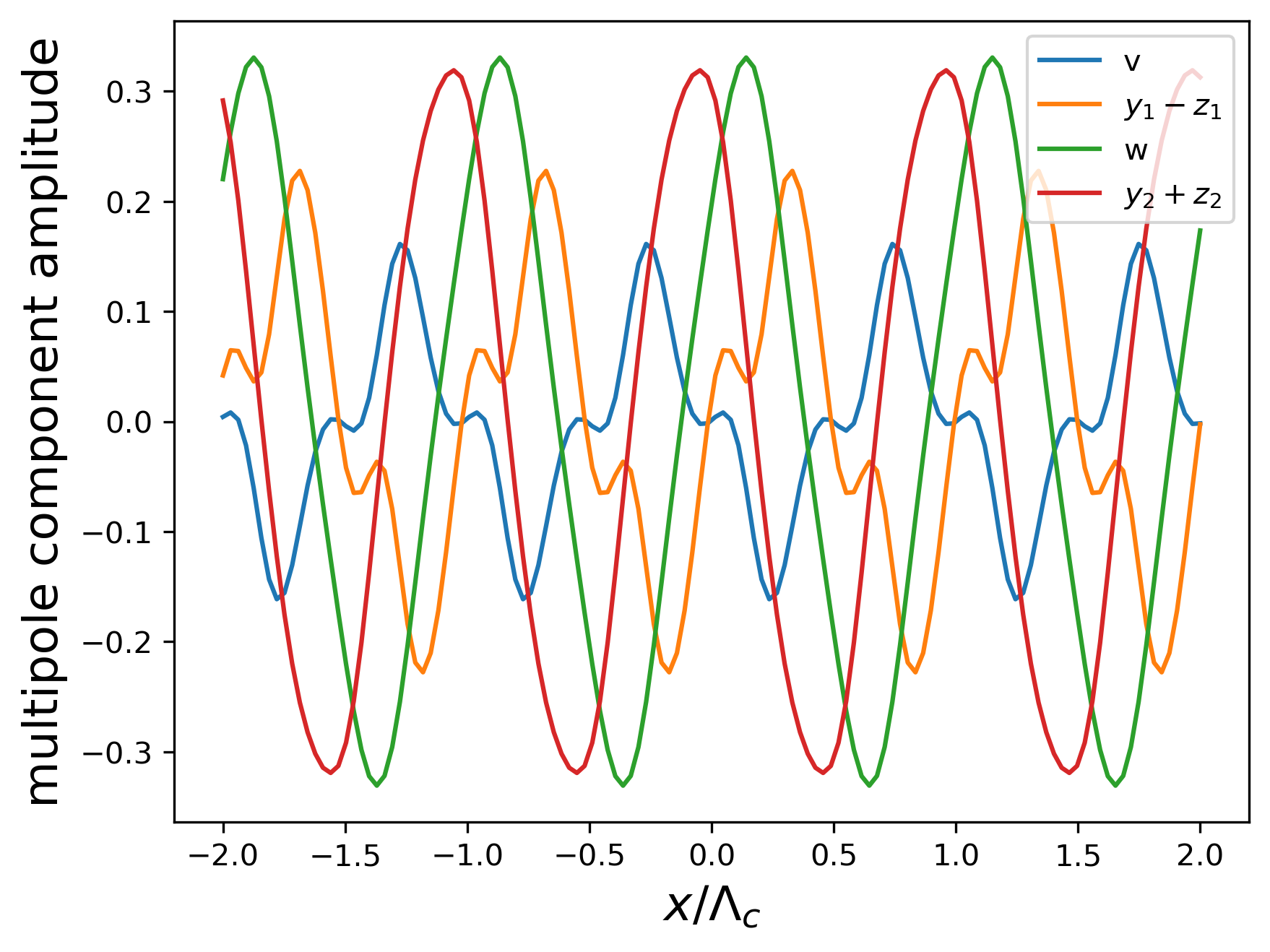}
\caption{Spatial profiles of snapshots of drifting structures. 
Upper panel  drifting left (anti-parallel to B-field), lower panel drifting right (in direction of B-field.  Parameters: $b_0$=70, pump intensity $I_{\rm{in}}=5$ mW/cm$^2$, $B_x=0.5$ G and detuning $\delta=-10\Gamma$. The graphs demonstrate the spatial sequence of the waves in the dipole ($w$, $y_2+z_2$) and quadrupolar components. Within the dipolar, respectively quadrupolar, components this is given by precession. The resulting screws are displayed in a 3D form in Fig.~2 of the main text. The locking between dipolar and quadrupolar is due to the mutually inhibitive coupling as discussed in text.}
\label{fig:stripes_cut}
\end{center}
\end{figure}

Fig.~\ref{fig:stripes_cut} shows spatial profiles of drifting structures. The profiles have a slight asymmetry; expressed to a different extent in different variables as expected for a structure with a spontaneous drift and hence parity breaking. For $B_x>0$ the spatial sequence of maxima in drift direction is \\
$v\xrightarrow[\Lambda/4]{} (y_2+z_2)\xrightarrow[\Lambda/4]{}w \xrightarrow[\Lambda/4]{} (y_1-z_1)\xrightarrow[\Lambda/4]{} v$,\\
 respectively\\
$m_x m_y + m_y m_x \xrightarrow[\Lambda/2]{}  m_z \xrightarrow[\Lambda/4]{} m_y,  m_x m_z + m_z m_x\xrightarrow[\Lambda/4]{} m_x m_y + m_y m_x $\\
for both drift directions. For $B_x<0$ it is \\
$v\xrightarrow[\Lambda/4]{} (y_1-z_1)\xrightarrow[\Lambda/4]{}w \xrightarrow[\Lambda/4]{} (y_2+z_2)\xrightarrow[\Lambda/4]{} v$,\\
 respectively\\
$m_x m_y + m_y m_x \xrightarrow[\Lambda/4]{}   m_y,  m_x m_z + m_z m_x\xrightarrow[\Lambda/4]{} m_z \xrightarrow[\Lambda/2]{} m_x m_y + m_y m_x $, \\
where $\Lambda/4$, $\Lambda/2$ indicate the spatial shift in units of the lattice period $\Lambda$.
This demonstrates that the sequence is dictated by the direction of precession; in line with the expectation from the analytical calculations in (\ref{eq:precession1_solution}) and (\ref{eq:precession2_solution_1}). Looking into the direction of propagation, both  $\vec{m}$ and $(v,y_2+z_2)$, respectively $(m_x m_y + m_y m_x,m_x m_z + m_z m_x)$, have right handed helicity in time at a specific location and form a left handed screw in space, if magnetic field and drift direction are parallel to each other, and a left handed helicity and a right handed screw,  if magnetic field and drift direction are anti-parallel to each other. (Although $m_x m_y + m_y m_x$, $m_x m_z + m_z m_x $ are formally not vector components, they will rotate around the $x$-axis similar to a vector: from being located within the $y$-plane into the $z$-plane and back into the $y$-plane with opposite sign.)

The synchronization between the dipole and quadrupole components is provided by the $\dot{v} = +P_{\Lambda +}/6 \, \Delta w + \ldots$ and $\dot{w} = -P_{\Lambda +}/6 \, \Delta v + \ldots$ terms in (\ref{eqs:transfields}) leading to anti-phase behaviour of $w$ and $v$, respectively $m_z$ and $m_x m_y + m_y m_x$ (Note that the dipole and quadrupole manifolds are not coupled by the magnetic field). This demonstrates that the drifting waves can form spontaneously with different handiness with respect to their propagation direction and hence a coupling between propagation direction and handiness or chiral behaviour.  We conjecture that the oscillatory coupling between dipolar and quadrupole components not possible in models with a $J=1/2$ ground state provides the destabilization of the stationary periodic state to a time dependent one, as numerical simulations indicate that the stationary state is destabilized if the Larmor frequency is of the order of or larger than the coupling provided by $\Delta P_{\Lambda +}/6$. Unfortunately, the theoretical model being based on eight coupled nonlinear partial differential equations is very complex and does not allow a straightforward analytical treatment.

\begin{figure}
\begin{center}
\includegraphics[width=\columnwidth,clip=true]{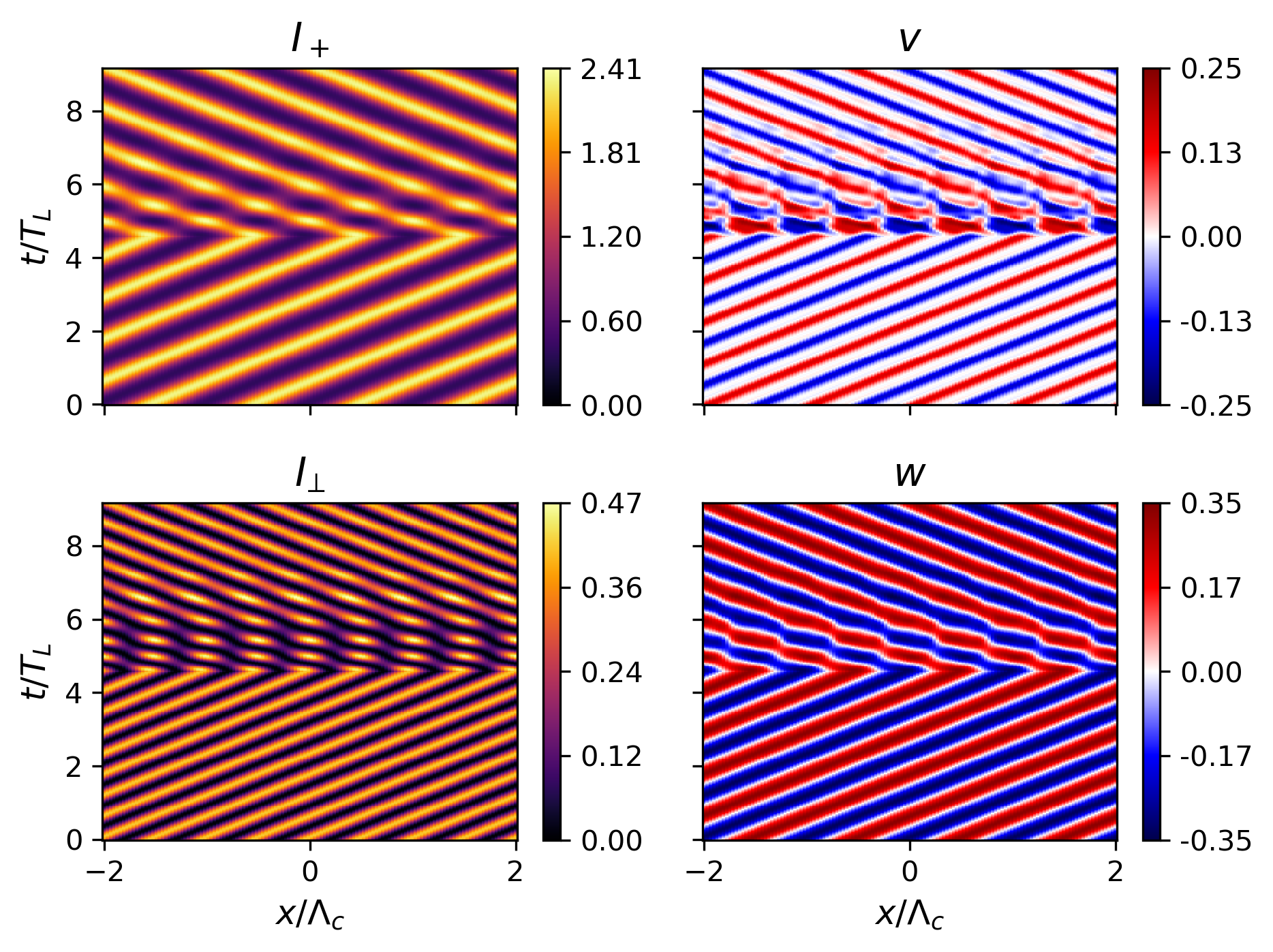}
\caption{Space time structure of drifting waves along $x$-axis for circular polarization component (upper left panel), linear polarization component perpendicular to input polarization (lower left panel), quadrupolar component $v$ (upper right panel), dipole component $w$ (lower right panel). The B-field is originally in positive $x$-direction and is flipped instantaneously to negative $x$-direction around time $t=5 T_L$. After a more complex transient, a `simple' wave structure is established again.
Parameters: $b_0$=70, pump intensity $I_{\rm{in}}=5$ mW/cm$^2$, $B_x=0.5$ G and detuning $\delta=-10\Gamma$.
}
\label{fig:flipBx-x_drift}
\end{center}
\end{figure}

For a given sign of the magnetic field, chirality and hence drift direction result from spontaneous symmetry breaking. However, if the sign of the magnetic field is flipped for a drifting structure (i.e.\ a structure with established chirality), the drift direction deterministically flips but the chirality is staying the same (Fig.~\ref{fig:flipBx-x_drift},  see also animation in \cite{labeyrie23s},
drift-x-flipBx-2D.mp4 for 2D pseudocolor structures and
drift-x-flipBx-screw.mp4 for vectorial reconstruction in `movies'). This indicates that chirality is the decisive degree of freedom originating from spontaneous symmetry breaking, not drift direction.

\begin{figure}
\begin{center}
\includegraphics[width=\columnwidth,clip=true]{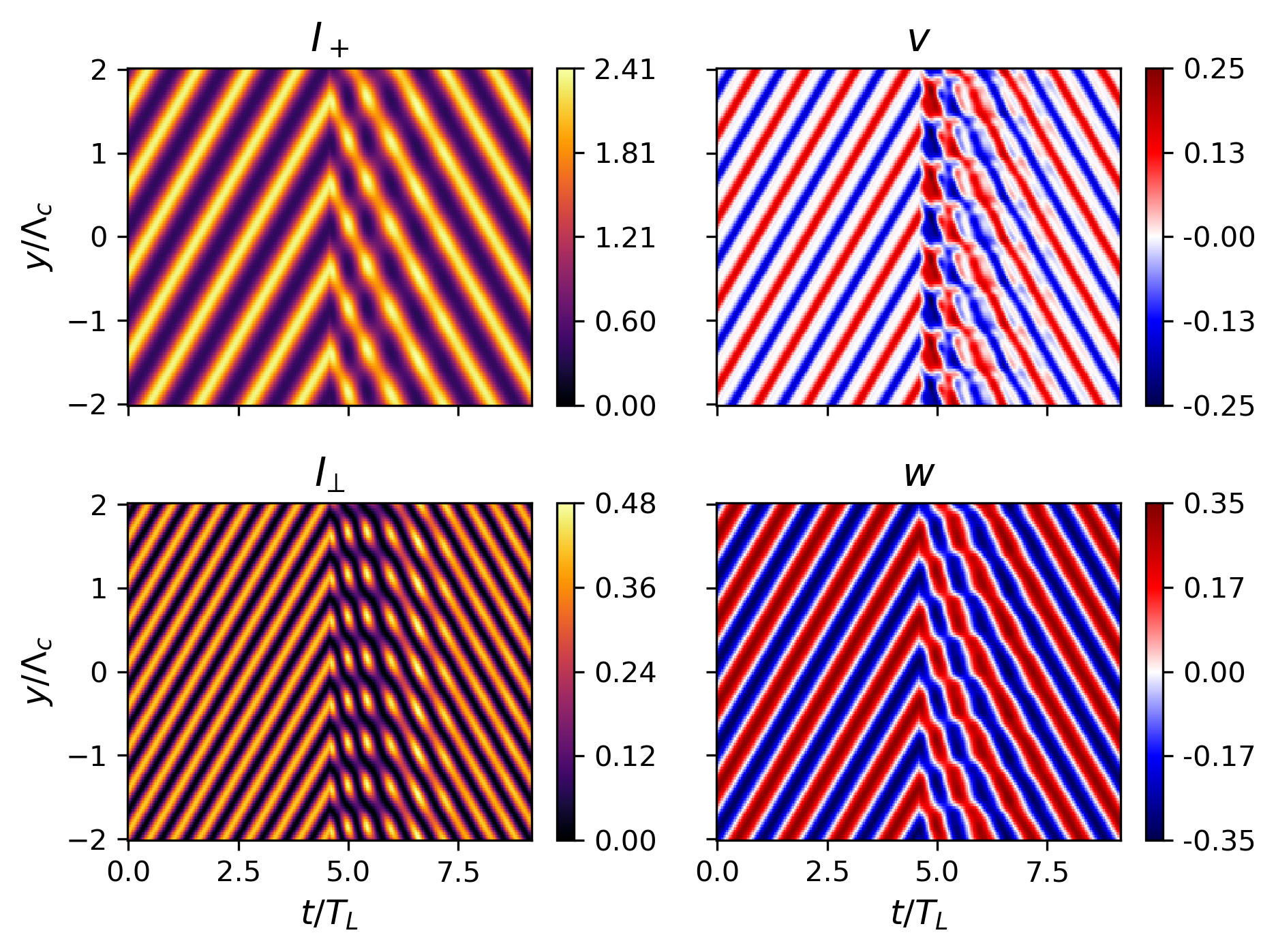}
\caption{Space time structure of drifting waves along $y$-axis for circular polarization component (upper left panel), linear polarization component perpendicular to input polarization (lower left panel), quadrupolar component $v$ (upper right panel), dipole component $w$ (lower right panel). The B-field is originally in positive $x$-direction and is flipped instantaneously to negative $x$-direction around time $t=5 T_L$. After a more complex transient, a `simple' wave structure is established again.
Parameters: $b_0$=70, pump intensity $I_{\rm{in}}=5$ mW/cm$^2$, $B_x=0.5$ G and detuning $\delta=-10\Gamma$.
}
\label{fig:flipBx-y_drift}
\end{center}
\end{figure}

Spontaneously drifting structures are also obtained if the Fourier filter is oriented along the $y$-axis, i.e.\ orthogonal to the applied magnetic field. The drift is visualized in the animation in \cite{labeyrie23s} (horizontal-rolls.mp4 in `movies'). Here the plane of precession is not orthogonal to the drift direction $x$ but in the plane spanned by the drift direction and the $z$-axis. Again, spontaneous symmetry breaking leads to chiral behaviour, as for a given sign of the magnetic field and hence direction of precession both drift directions are possible.  This corresponds to the situation in chiral quantum optics \cite{lodahl17} where the longitudinal polarization component of a strongly non-paraxial light fields and one transverse one couple to elliptically or ideally circularly polarized light with the sense of rotation depending on propagation direction. Also here, switching the sign of the magnetic field for an established drifting structure flips the drift direction but preserves chirality (Fig.~\ref{fig:flipBx-y_drift}).

\end{document}